\documentclass[aps,singlecolumn,superscriptaddress]{revtex4-2}
\usepackage{amsmath,amssymb}
\usepackage{graphics,graphicx}
\usepackage{subfigure}
\usepackage{dcolumn,bm}
\usepackage{psfrag}
\usepackage{color}
\usepackage{natbib,hyperref}
\usepackage{multirow}
\usepackage{tabularx}
\usepackage{braket}
\usepackage{orcidlink}
\newcommand{\vc}
{\affiliation{Vidyasagar College, 39 Sankar Ghosh lane, Kolkata 700006, India.}}
\newcommand{\be}
{\begin{equation}}
	\newcommand{\ee}
	{\end{equation}}

\begin{document}
	\title{An Insight of Heart-Like Systems with Percolation}
	\author{Md Aquib Molla \orcidlink{0000-0003-0416-1349}}
	\vc
	\author{Sanchari Goswami \orcidlink{0000-0002-4222-5123}}
	\vc
	
	\begin{abstract}
		We study the signal percolation through heart-like biological system. Starting from an initial distribution of waiting and inactive cells with probabilities $p$ and $(1-p)$ respectively, the signal propagation is observed in terms of active cells. As the signal enters the system from one end, the number of arrival of active sites at the other end is studied and analysis of the system behaviour is made by varying a few important parameters of the system like $p_{switch}$ (switching probability from inactive to waiting) and $p_{act}$ (switching probability from waiting to active). In this connection, the non-regular heart rhythms are discussed. Fraction of paths percolating through the system shows a transition from $0$ to $1$ near $p=p_c$. Some other important quantities like tortuosity and cluster distribution are discussed. Several critical exponents have been obtained and compared the exponents of standard percolation. 
	\end{abstract}
	
	\maketitle
	
	\section*{Introduction}
	Percolation is one of the simplest and fundamental models in statistical mechanics which is not always an exactly solvable model. In spite of its simple update rules, it can be applied to describe a wide variety of systems, e.g., fluid flow through porous media \cite{Hammersley}, electrical movement in neurons, fibrosis in organs, damage spreading in solids, disease spreading in a community, opinion formation in society or even for knowledge percolation \cite{dob, barghathi, Vigmond, Niederer, rabi, soumya, anjan, bagnoli}. A very popular model of percolation is the forest fire model (FFM) introduced in \cite{Bak, stauffer, Song} mainly as a 2D lattice studied as ``directed percolation" (DP) \cite{drossel, drossel2},  isotropic percolation (IP) \cite{stauffer} and semi directed percolation (SDP) \cite{SDP, redner, Martin} in several variations. 
	
	The movement of signal in neurons for the heart and brain systems \cite{dob, barghathi, Vigmond, Niederer} can be studied in the light of percolation. Movement of heat and electrical signal through different media have been studied in \cite{langlois, gibson} by percolation methods. In addition to this, there are certain important quantities like `Tortuosity' which have been studied for a few systems in combination with percolation \cite{ghanbarian}. 
	
	This process of signal percolation through heart is an interesting topic to study. When the heartbeats do not follow the usual pattern but become very fast or irregular, it leads to a disease called arrhythmia \cite{Desai}. Arrhythmias in turn cause health situations such as fainting, stroke, heart attack or even death. Normally, a special group of cells in the sinoatrial (SA) node in the upper right chamber of the heart begin the electrical signal to start the heartbeat. This signal travels down the heart towards the ventricles, i.e., the two lower chambers of the heart. This organized pattern helps the heart to beat in its usual way. However, several problems affect the normal conduction of signal and lead to abnormal heart rhythms. Due to faster beats, there may not be enough time in between beats and this may prevent the heart from pumping the required amount of blood to the whole body which may lead to arrhythmia and related health problems. 
	
	Several biological models have been used to study the heart system with the help of computation. Computational models in cardiology showed enormous scope in revealing diagnostic information related to cardiac disease. At the same time there are provisions that these models help the diagnosis to be cost-effective and to have low risk \cite{Niederer}. The cells can be activated by an external stimulus which initiates an action potential (AP). The cells after activation will go through a certain refractory period. The exact physiological processes are described in \cite{Dossel} in a detailed manner. In \cite{rabi}, the model of heart is studied with three states of cells: active, inactive and waiting in connection with the AP. 
	
	One of the simplified system to study a percolation problem is a $L \times L$ lattice, where there is a parameter called percolation probability $p$ with which a site could be open, i.e. can percolate ``something". This in turn indicates that a site is closed with $1-p$ probability, means unable to percolate. It is well known from different studies that below a critical percolation probability $p_c$ the system is unable to percolate, whereas above $p_c$ infinite percolation cluster can be formed \cite{stauffer,Derrida,Malarz,Malarz2,Feng,Mertens}. In connection to percolation there are a number of exponents which can be studied as well. Those exponents actually describes the system near criticality \cite{stauffer, Kopelman,Mario}.
	
	The heart-like system is studied here in the light of percolation theory. In this study, we tried to discuss several features related to the AP movement as in IP through the heart-like system by varying a few parameters in connection to the heart cells.  Several critical exponents for the percolation model of the heart-like system are obtained and compared to the corresponding exponents obtained from standard percolation.

	\section*{Description of the Model}
	We study a 2D square grid of $L \times L$ cells where initially the cells may be in two states, ``waiting" and ``inactive" as described in \cite{rabi} also. A ``waiting" cell is an open cell which is ready to hold anything percolating through the system. An ``inactive" cell is however equivalent to a closed cell which is unable to percolate. We choose the system with an initial distribution of ``waiting" cells with probability $p$ and ``inactive" cells with probability $1-p$. A ``waiting" cell can be activated with an AP (the ``thing" that is actually percolating through the system) and then it becomes an ``active" one. The condition for a waiting cell to transform into an active one in the next step is that one or more of its ``nearest cell neighbors" should be active. An ``active" cell which is holding the AP at a particular instant will be transformed into an ``inactive" one in the next time step. For a heart system this is consistent with the fact that there is a certain refractory period of the cells.
	
	From now on, we will use numbers to indicate the type of the cells for our convenience. Thus
	\begin{itemize}
		\item an ``waiting" cell will be identified as ``$0$", 
		\item an ``active" cell will be identified as ``$1$",
		\item an ``inactive" cell will be identified as ``$2$".
	\end{itemize}
	
	The general rule of transformation for a cell may be described as: waiting $\rightarrow$ active $\rightarrow$ inactive, or, using the numbers as $0 \rightarrow 1 \rightarrow 2$. However, in some specific cases an ``inactive" cell may become a ``waiting" one, i.e., $2 \rightarrow 0$ is also allowed. This is similar to the tree regrowth at several sites in FFM as discussed in \cite{Bak, stauffer}.
	
	In addition to these rules, some definite probabilities are considered to be associated with this system. We introduce two probabilities:
	\begin{itemize}
		\item $p_{act}$(inhibitory) by which an waiting cell can become an active one, i.e., $0 \rightarrow 1$, 
		\item $p_{switch}$(refractory) by which an inactive cell can be reverted to a waiting one, i.e., $2 \rightarrow 0$. 
	\end{itemize}.
	
	The AP is initiated from one side (here from the top) of the 2D matrix. This is done by labelling all the cells on the first row as $1$. The different rows are updated at different times $t$. It is obvious that to receive the AP at the bottom/final row the number of time steps required will at least be equal to the system size $L$. 
	
	The system is studied from two aspects. First, in subsection A of the Results, the number of arrival $N_A$ at the other end of the 2D matrix is observed with variation of the parameters $p_{act}$ and $p_{switch}$. If there is no inactive cell, the AP reach all the cells of the final row in time $L$. However, when there is a distribution of $0$ and $2$ cells, all the cells of the final row do not receive the AP at the same time but $1$s arrive at different cells at different times. At the final row the number of cells containing the AP is counted at each time. Second, in subsection B and C of the Results, the percolation model of the system is analyzed. The delay in arrival is measured and some interesting properties related to that are studied. In the same context the cluster distribution and several critical exponents related to the system are also studied.
	
	\section*{Results}
	\subsection{Study of Number of Arrival}
	The AP, as initiated from one end of the 2D lattice of size $L$, percolates through the system. The number of arrivals at the other end, i.e., the number of active sites $N_A$ at that end is checked with time $t$ . We study the number of arrivals $N_A$ with two conditions:
	\begin{enumerate}
		\item When the $2 \rightarrow 0$ transition is not allowed $p_{switch}=0.0$ : This arises when the cells after becoming active receiving the AP for once become inactive and stay in their refractory period for a very long time. Here $p_{switch}=0.0$ is equivalent to an infinite refractory period. This can be often seen due to certain health condition. For old people this is very common \cite{shirakabe}. At the same time it may also happen that the transition from $0 \rightarrow 1$ can occur with probability less than $1$. This may also happen due to some irregularity in the cell activation \cite{anversa}.
		\item When $2 \rightarrow 0$ transition is allowed with a certain probability other than $0.0$ :  Here $p_{switch} \neq 0.0$ indicates that the cells are in their refractory period. However, very small value of $p_{switch}$ indicates that the cells are in their refractory period for longer than usual. Medications and better lifestyle may lead the system to have a higher $p_{switch}$. However, having a very larger value of $p_{switch}$ is not suitable for this case and will be discussed later.
	\end{enumerate}
	
	In addition to $p_{switch}=0.0$ there may be variation of $p_{act}$ with which we have studied the system.
	
\begin{figure}[h!]
		\begin{center}
			\hspace*{-0.1in}
			\includegraphics[angle=-90, trim = 0 0 0 0, clip = true, width=0.75\linewidth]{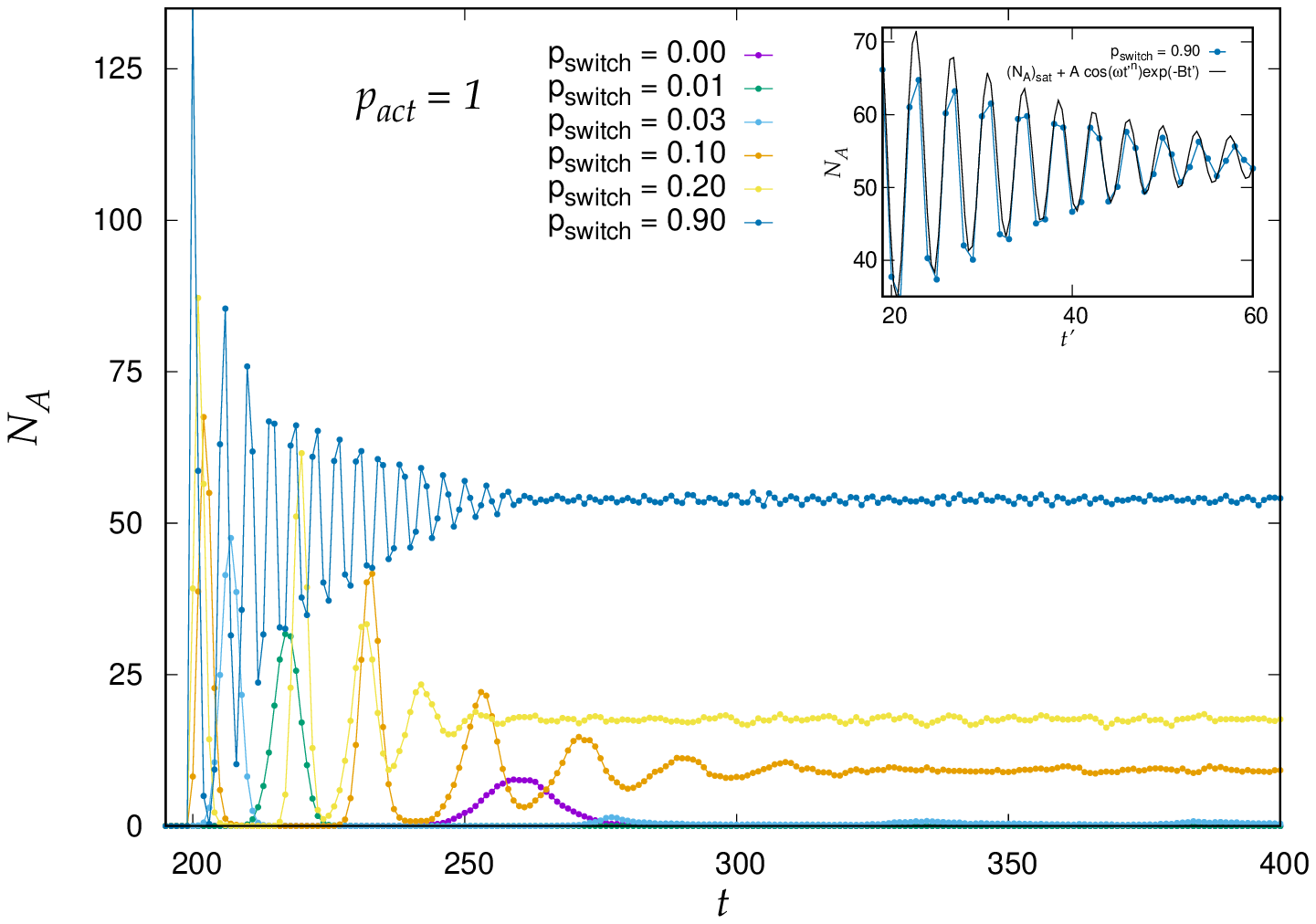}
			\caption{$N_{A}$ versus $t$ with $L$ = 200, $p = 0.70$, $p_{act} = 1.0$ and $p_{switch}$ varying from $0$ to $0.90$. In the inset $p_{switch} = 0.90$ has been shown with equation \ref{Damped} where $(N_A)_{sat} \simeq 54$, $A = 55$ and $B = 0.05$.}
			\label{Varying Pswitch}
		\end{center}
	\end{figure}

	The variation of number of arrival $N_A$ with $t$ is shown in Fig. \ref{Varying Pswitch} for $p_{act} = 1$ and $p_{switch}$ varying from $0.0$ to $0.9$. The plots show some kind of oscillatory behaviour for high values of $p_{switch}$ and finally saturate to some specific non-zero values $(N_A)_{sat}$. 
	The oscillatory behaviour can be explained in the following way. It is well known that when a fluid  is flowing through a porous medium, there is some kind of relative motion \cite{Johnson} which generates a fluid shear \cite{langlois}, can be considered as a friction/damping. In the same way, when an electrical signal, here the AP, is moving through the heart, there must be some retardation present. For high $p_{switch}$, the motion is similar to an underdamped oscillator. This is obvious as for high $p_{switch}$, the signal more frequently finds a path to percolate, be it in forward or backward directions or sideways. Therefore number of arrival $N_A$ is never zero. However, for $p_{switch}=0.0$, it has been observed that the behaviour of $N_A$ is Gaussian. This signifies that for low $p_{switch}$ the number of active cells die down almost completely in time $L<t<2L$. Therefore, there is no chance for further AP signals to pass down through the system. However, for higher $p_{switch}$, this is not true and even after a very long time, there is always a chance for the next APs to pass through the system. This is because here we have $(N_{A})_{sat} > 0$ always. It indicates that, a certain finite number of active cells will always be present in the system. Even if we study the system for $t \rightarrow \infty$, the system still shows non-zero active sites. \\
	
	It is already observed that for $p_{switch} \rightarrow 1$ the system shows a damped oscillatory behaviour. 
	Therefore it is worth comparing the system parameters to the parameters associated with a damped oscillator. 
	The differential equation of a damped oscillator is 
	\begin{equation}
		\frac{d^2x}{dt^2}+2B\frac{dx}{dt}+\omega_0^2x = 0,
	\end{equation}
	where the coefficient $2B$ is related to damping and $\omega_0$ is the natural frequency of oscillation. For underdamped/oscillatory motion $B^2 < \omega_0^2$ and the corresponding solution can be written as,
	\begin{equation}
		x(t) = A \cos(\omega t) \exp(-Bt),
	\end{equation}
	where $\omega = \sqrt{\omega_0^2-B^2}$.
	For this purpose, we fit $N_{A}$ with the following function:
	\begin{equation} \label{Damped}
		N_{A}(t) = (N_A)_{sat} + A \cos(\omega {t'}^{\nu}) \exp(-Bt'),
	\end{equation}
	where $t' = t - L $. The approximate values are shown in Table \ref{Dampped oscillator}.
	
	
	\begin{table}[h!]
		\begin{center}
			\begin{tabular}{|c|c|c|c|c|c|}
				\hline
				\hspace{0.4cm}$p_{switch}$\hspace{0.4cm} & \hspace{0.6cm}$(N_A)_{sat}$\hspace{0.4cm} & \hspace{0.4cm}$A$\hspace{0.4cm} & \hspace{0.4cm}$\omega$\hspace{0.4cm} & \hspace{0.4cm}$\nu$\hspace{0.4cm} & \hspace{0.4cm}$B$\hspace{0.4cm} \\  
				\hline
				\hline
				0.90	&	53.83    &	55 & 1.00 & 1.10 & 0.05\\
				\hline
				0.80	&	48.78    & 58 & 0.65 & 1.20 & 0.07\\
				\hline
				0.70	&	45.31    &	70 & 0.45 & 1.29 & 0.09\\
				\hline
				0.60	&	41.21    &	75 & 0.31 & 1.37 & 0.10\\
				\hline
			\end{tabular}
		\end{center}
		\caption{Approximate values of the parameters in Equation \ref{Damped} with $L = 200$.}
		\label{Dampped oscillator}
	\end{table}
	
	In case of $p_{act} = p_{switch} = 1$ (not shown in Fig. \ref{Varying Pswitch}) the oscillatory behaviour continues as the rows are updating as waiting $\rightarrow$ active $\rightarrow$ inactive $\rightarrow$ waiting.
	
\begin{figure}[h!]
		\begin{center}
			\hspace*{-0.1in}
			\includegraphics[angle=-90, trim = 0 0 0 0, clip = true, width=0.75\linewidth]{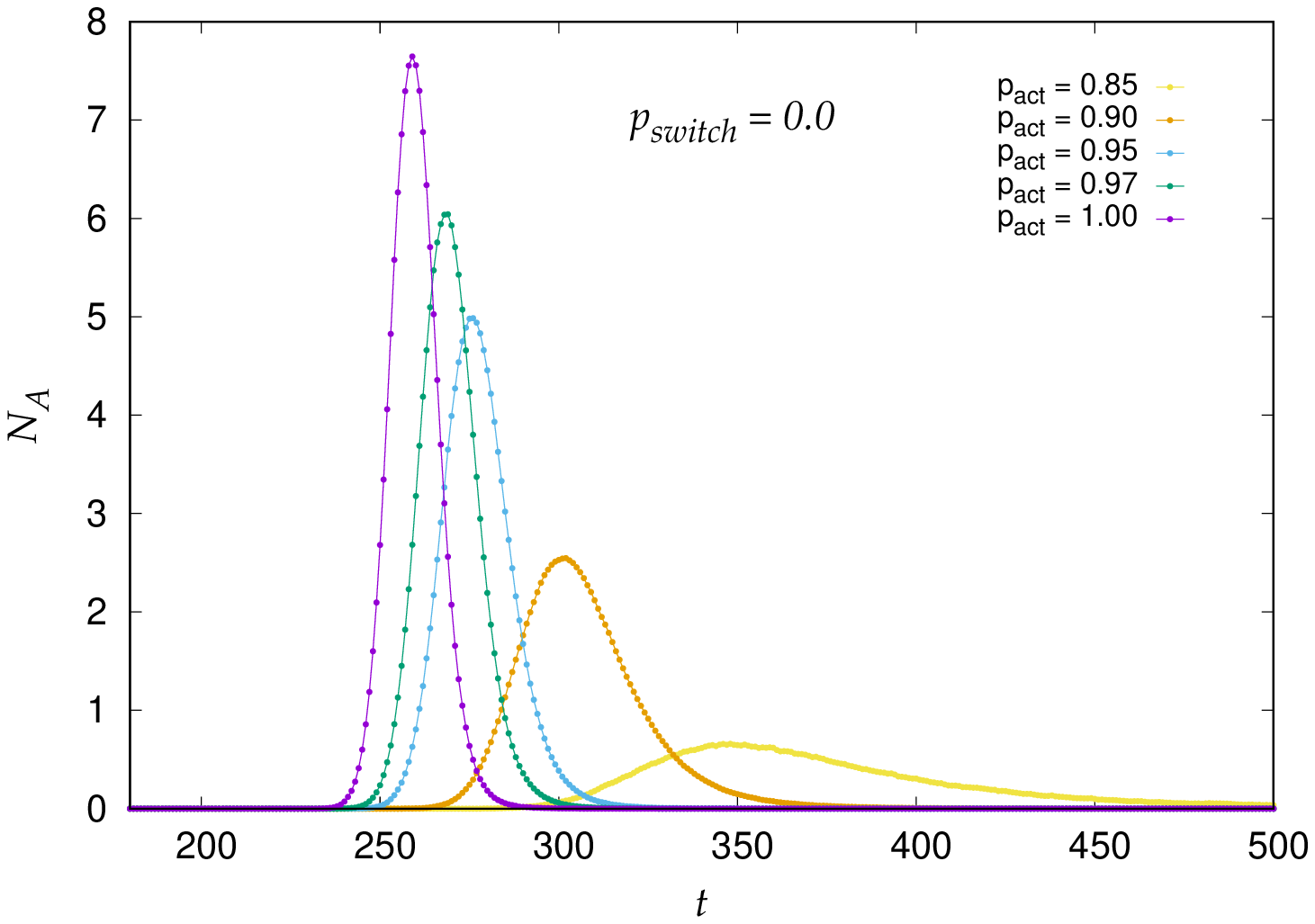}
			\caption{$N_{A}$ versus $t$ with $L = 200$, $p_{switch} = 0.0$ and $p_{act}$ varies from $0.85$ to $1.00$.}
			\label{Varying Pact}
		\end{center}
	\end{figure}

	We now fix $p_{switch} = 0$ and study the system behaviour by varying $p_{act}$. It is clear from Fig. \ref{Varying Pswitch} that the behaviour will be approximately Gaussian. In Fig. \ref{Varying Pact} the behaviour is shown for $0.85 \leqslant p_{act} \leqslant 1.0$. The parameters related to the Gaussian behaviour is shown in Table \ref{Table Varying Pact} for $L=200$ by fitting with the following function:
	\begin{equation} \label{Gaussian}
		N_{A}(t) = (N_A)_{max} e^{-\frac{1}{2} \left( \frac{t-\mu}{\sigma_{sd}} \right)^2},
	\end{equation}
	where $(N_A)_{max}$ is the peak value, $\mu$ is the mean and $\sigma_{sd}$ is the standard deviation of the Gaussian profile.
	
	\begin{table}[h!]
		\begin{center}
			\begin{tabular}{|c|c|c|c|}
				\hline
				\hspace{0.2cm}$p_{act}$\hspace{0.2cm} & \hspace{0.2cm}$(N_A)_{max}$\hspace{0.2cm} & \hspace{0.6cm}$\mu$\hspace{0.6cm} & \hspace{0.4cm}$\sigma_{sd}$\hspace{0.4cm} \\  
				\hline
				\hline
				$1.00$	&	$7.78$  &	$259.10$    &     $6.35$\\
				\hline
				$0.97$	&	$6.20$    &	$268.32$     &     $7.48$\\
				\hline
				$0.95$	&	$4.89$    &	$275.54$     &    $9.22$\\
				\hline
				$0.90$	&	$2.66$    &	$303.00$     &    $14.01$\\
				\hline
				$0.85$	&	$0.66$    &	$356.50$     &    $33.71$\\
				\hline
			\end{tabular}
		\end{center}
		\caption{Approximate values of the parameters in Equation \ref{Gaussian} with $L = 200$.}
		\label{Table Varying Pact}
	\end{table}
	
\begin{figure}[h!]
		\begin{center}
			\hspace*{-0.1in}
			\includegraphics[angle=-90, trim = 250 0 0 0, clip = true, width=0.99\linewidth]{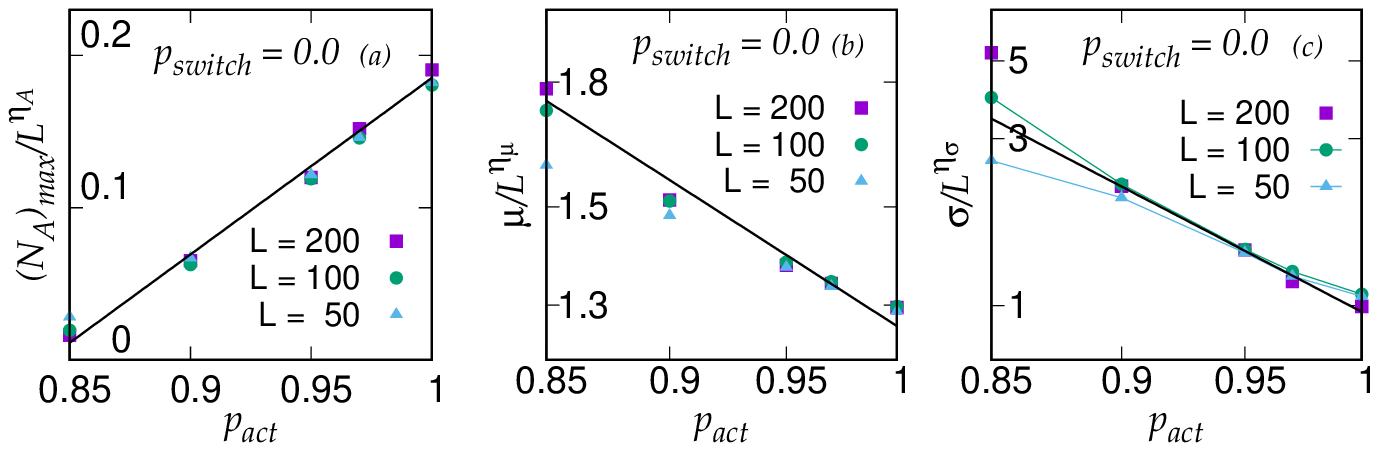}
			\caption{Variations of the Gaussian parameters with $p_{act}$ where $L = 50, 100$ and $200$, (a) $\frac{(N_{A})_{max}}{L^{\eta_{A}}}$ versus $p_{act}$ with $\eta_{A} = 0.7$. For the fitted straight line, slope $m = 1.161 \pm 0.055$ and intercept $c = -0.976 \pm 0.051$,
				(b) $\frac{\mu}{L^{\eta_{\mu}}}$ versus $p_{act}$ (in log scale) and the fitted function is $a_{\mu} p_{act}^{b_{\mu}}$ with $\eta_{\mu} = 1.0$, $a_{\mu} = 1.261 \pm 0.026$ and $b_{\mu} = -2.023 \pm 0.201$ 
				and (c) $\frac{\sigma}{L^{\eta_{\sigma}}}$ versus $p_{act}$ (in log scale) and the fitted function is $a_{\sigma} p_{act}^{b_{\sigma}}$ with $\eta_{\sigma} = 1.0$, $a_{\sigma} = 0.960 \pm 0.028$ and $b_{\sigma} = -7.819 \pm 0.358$. 
			}
			\label{Gaussian parameters}
		\end{center}
	\end{figure}

	It is observed that as $p_{act}$ decreases, $(N_A)_{max}$ decreases. It is in agreement to our description of the model. For $p_{act}<1$ the waiting cells do not receive the AP with $100\%$ probability and therefore cannot become active. Therefore $(N_A)_{max}$ decreases. The movement of the peak towards right with decreasing $p_{act}$, i.e., shifting of the mean of the Gaussian profile towards higher $t$ can also be explained in the following manner: As $p_{act}<1$ the signal cannot find its way smoothly inside the system and therefore must take a longer time to appear at the other end. The variations of $(N_A)_{max}$, $\mu$ and $\sigma_{sd}$ are shown as a function of $p_{act}$ in Fig, \ref{Gaussian parameters} for $L = 50, 100, 200$.  
	From Fig. \ref{Varying Pact} it is evident that with decrease of $p_{act}$ the curve broadens and the height of the peak decreases. This suggests that not only arrivals are delayed because of obstacles within the system, the signal is getting weakened too. If we analyze the area under the curve carefully, it is evident that it is decreasing with decrease in $p_{act}$. This also confirms presence of some kind of retardation as it usually present for a real heart system.\\
	
	\subsection{Study of Percolating Paths and Tortuosity}
\begin{figure}[h!]
		\begin{center}
			\hspace*{-0.1in}
			\includegraphics[angle=-90, trim = 0 0 0 0, clip = true, width=0.75\linewidth]{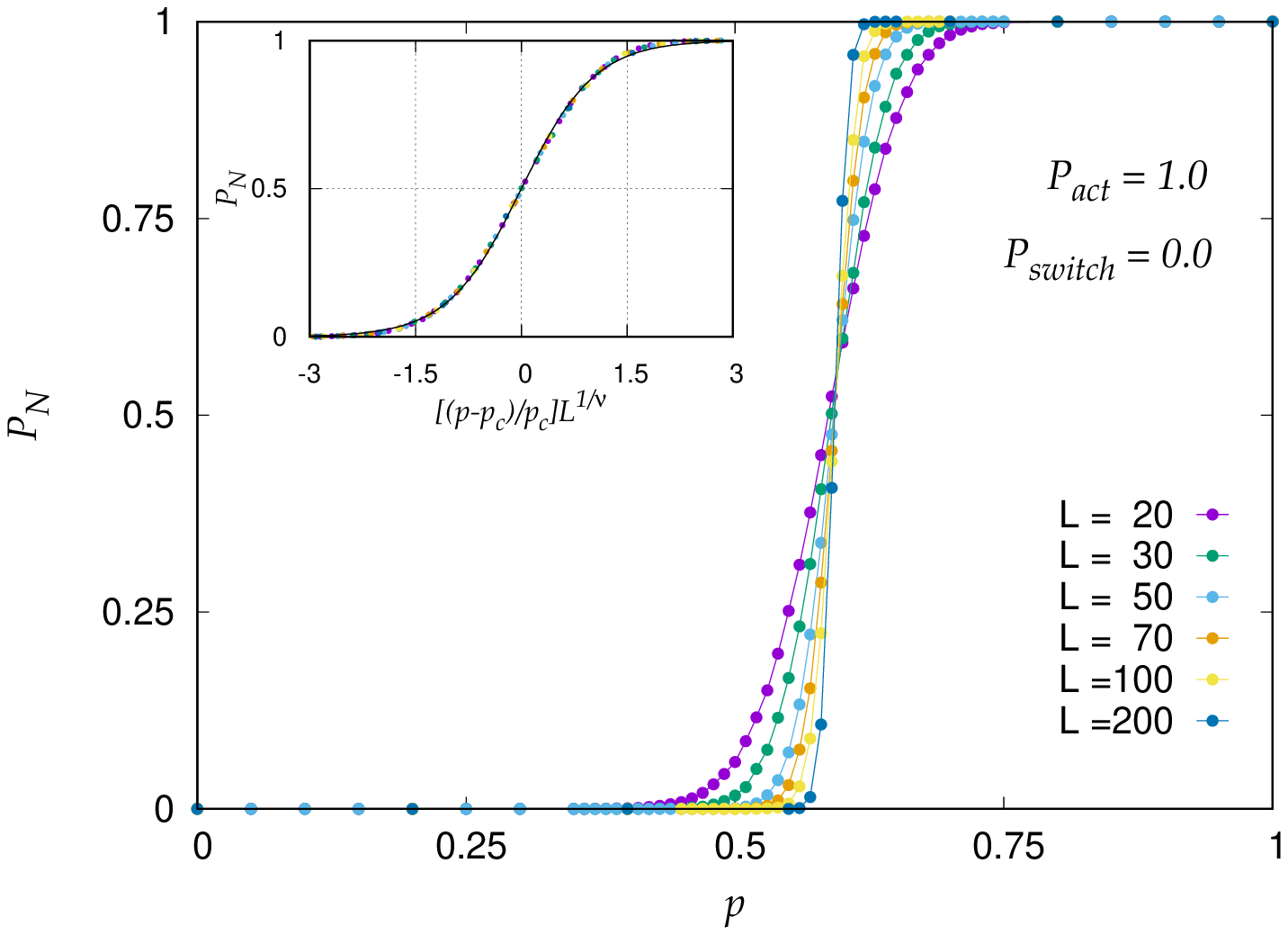}
			\caption{$P_N$ versus $p$ with $p_{act} = 1.0$, $p_{switch} = 0.00$ where $L =  20, 30, 50, 70, 100$ and $200$ in the inset data collapse has been shown as $\frac{(1+\tanh(\kappa x))}{2}$ with $\nu = \frac{4}{3}$ and $\kappa = 0.96$.}
			\label{Percolation_Pact1_Pswitch0}
		\end{center}
	\end{figure}

\begin{figure}[h!]
		\begin{center}
			\hspace*{-0.1in}
			\includegraphics[angle=-90, trim = 150 0 0 0, clip = true, width=0.85\linewidth]{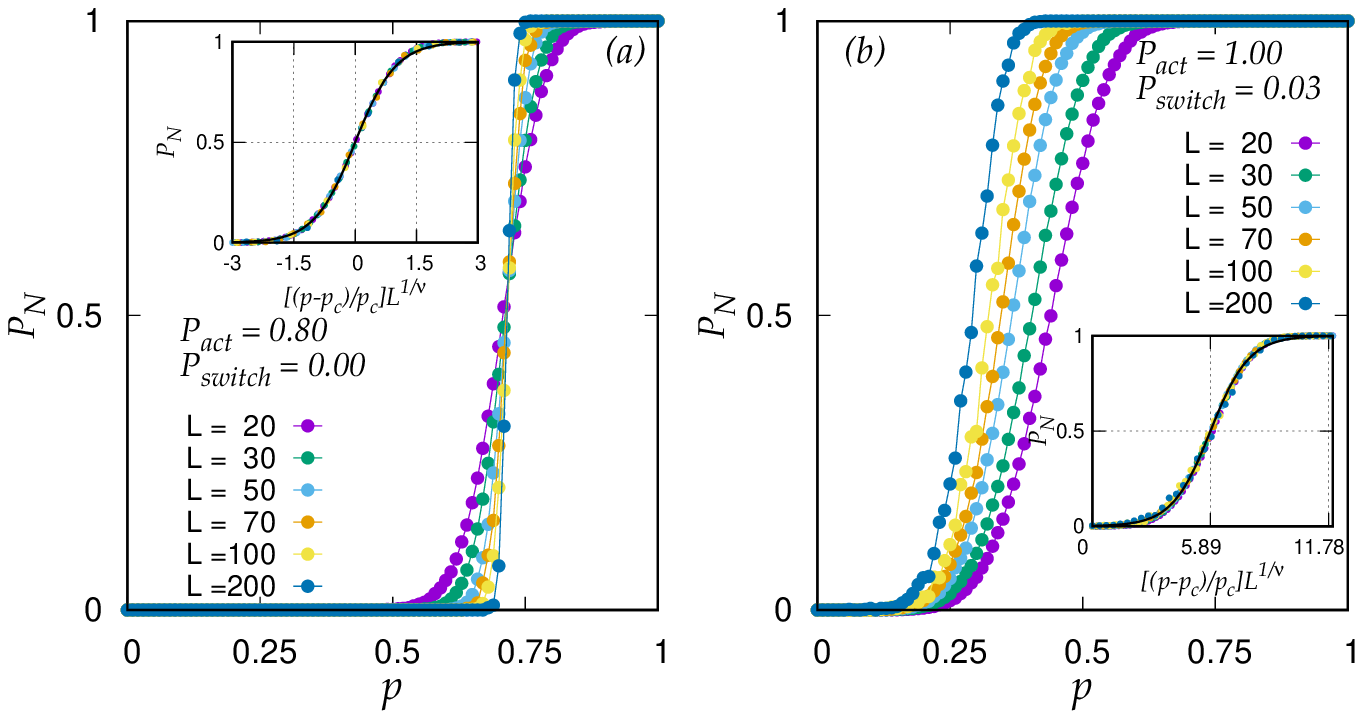}
			\caption{$P_N$ versus $p$ where $L =  20, 30, 50, 70, 100$ and $200$, $(a)$ $p_{act} = 0.80$ and $p_{switch} = 0.00$. In the inset data collapse has been shown and the fitted function is $\frac{(1+\tanh(\kappa x))}{2}$, $\nu = \frac{4}{3}$ and $\kappa = 1.0$ $(b)$ $p_{act} = 1.0$ and $p_{switch} = 0.03$. In the inset data collapse has been shown and the fitted function is $\frac{(1+\tanh(\kappa (x-\epsilon))}{2}$, $\nu = 3.7$, $\kappa = 0.52$ and $\epsilon = 5.89$.}
			\label{Percolation_collage}
		\end{center}
	\end{figure}

	Next we study the fraction of percolating paths out of $N_T$ number of trials, termed as probability $P_N$ here, as a function of $p$. This is shown in Fig. \ref{Percolation_Pact1_Pswitch0} for $p_{act}=1.0, p_{switch}=0.0$ for different $L$. Here we have taken $N_T=10^5$. In the inset of Fig. \ref{Percolation_Pact1_Pswitch0} the data collapse has been shown. It has been observed that the percolation threshold is at $p_c = 0.5925 \pm 0.0001$ which is in good agreement with \cite{Derrida,Malarz,Feng,Mertens}. The function is checked to be fitted with the function $\frac{(1+\tanh{\kappa x})}{2}$ where $\kappa = 0.96$, $x=(\frac{p-p_c}{p_c})L^{1/{\nu}}$ and $\nu=\frac{4}{3}$. The same has also been studied for other combinations of $p_{act}$ and $p_{switch}$. As we decrease the value of $p_{act}$, the percolation threshold $p_c$ shifts towards higher side, as the waiting sites are becoming active with less probability. If keeping $p_{act}=1.0$ we increase the value of $p_{switch}$ even slightly, $p_c$ shifts towards lower $p$. This is also in agreement with Fig. \ref{Varying Pswitch}. These are shown in Fig.\ref{Percolation_collage} (a) for $p_{act} = 0.80$, $p_{switch} = 0.00$ and (b) $p_{act}=1.0$, $p_{switch} = 0.03$. From Fig. \ref{Percolation_collage}(a) $p_c$ is found to be $0.7157$ and from \ref{Percolation_collage}(b) it is found to be close to $0.12$. It is to be noted that in Fig. \ref{Percolation_collage}(b) there is no crossing and even for very small $p_{switch}$, $p_c$ drops rapidly. 
	
	It is interesting to study the delay of arrival of the signal at the receiving end by analyzing the paths of the active cells for the cases when $p_{switch}=0.0$. For this we have calculated the \textit{Tortuosity} $\tau$  \cite{Niederer}, which is defined as,
	\begin{equation} \label{tor}
		\tau = \frac{\braket{l_0}}{L} = \frac{x_0}{L},
	\end{equation}
	where $\braket{l_0}=x_0$ is the average path length.
	
\begin{figure}[h!]
		\begin{center}
			\hspace*{-0.1in}
			\includegraphics[angle=-90, trim = 0 0 0 0, clip = true, width=0.75\linewidth]{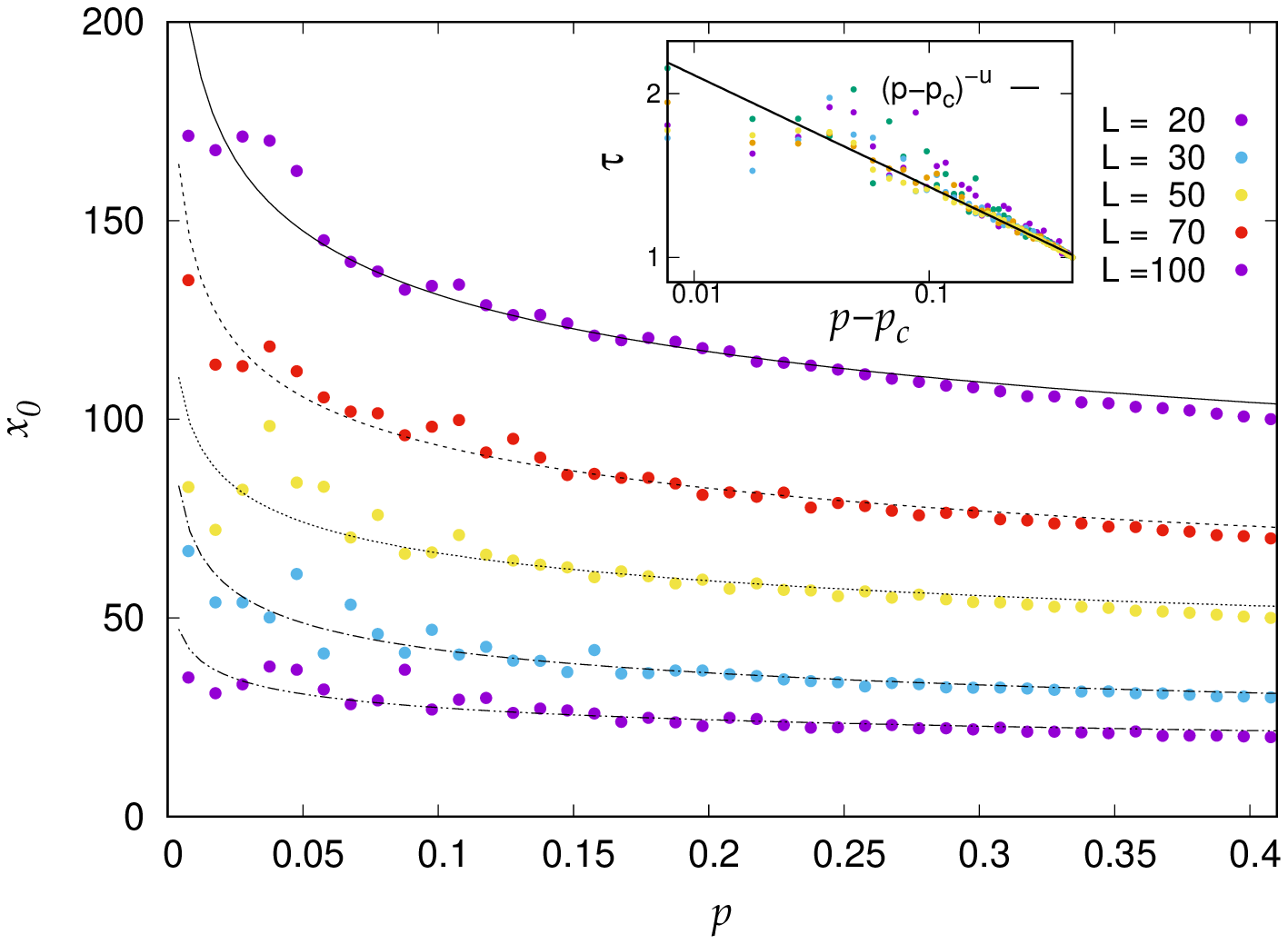}
			\caption{$x_0$ versus $p$ with $p_{act} = 1.00$ and $p_{switch} = 0.0$ and $L = 20, 30, 50, 70$ and $100$ n the inset data collapse has been shown as $\tau$ versus $p$ with $u = 0.205 \pm 0.003$.}
			\label{Tortuo}
		\end{center}
	\end{figure}

	Of course, $\tau$ depends on percolation probability $p$, since below critical percolation $p_c$ signal cannot be transmitted through the system. The variation of $x_0$ as a function of $p$ is shown in Fig. \ref{Tortuo} and $\tau$ as a function of $p$ is shown in the inset of the same Figure in log scale for $L = 20, 30, 50, 70$ and $100$. It has been observed that,
	\begin{equation}
		\tau \propto (p-p_c)^{-u}.
	\end{equation}
	The exponent $u$ is almost independent of $L$ and the approximated value is $u \simeq 0.205 \pm 0.003$. It is to be mentioned here that value of this exponent remains unaltered with variation of $p_{act}$.  
	
	\subsection{Study of Cluster Distribution and Critical Exponents}
	
	Now we would move on to find out some other relevant exponents for the system. It is well known that, the cluster distribution for site percolation shows power law behaviour for $p=p_c$ as $n(s)_{p_c} \propto s^{-{\Delta}}$, where $\Delta$ is the Fisher exponent \cite{stauffer}. 
	\begin{figure}[h!]
		\begin{center}
			\hspace*{-0.1in}
			\includegraphics[angle=-90, trim = 0 0 0 0, clip = true, width=0.75\linewidth]{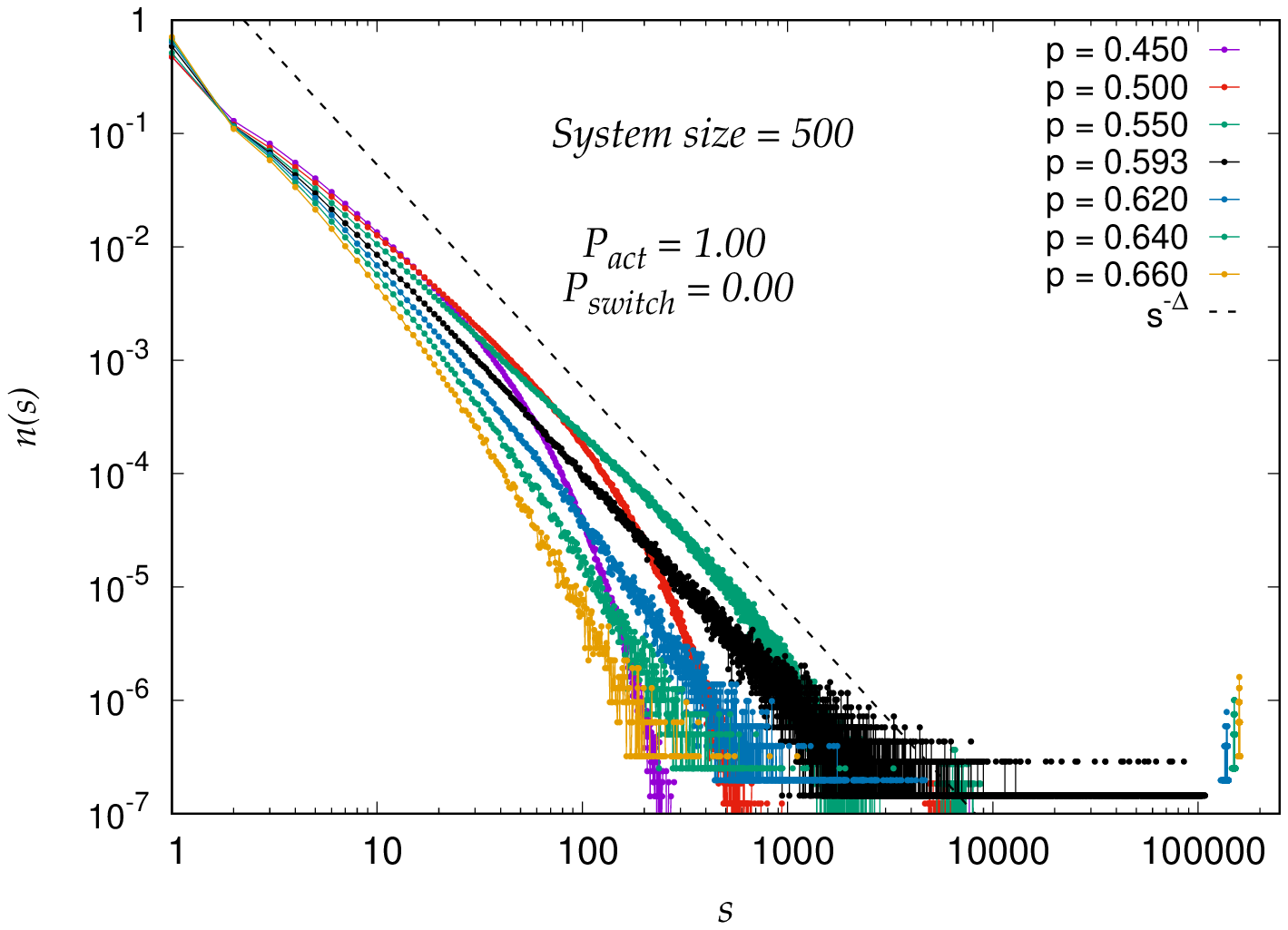}
			\caption{$n(s)$ versus $s$ with $L = 500$, $p_{act} = 1.00$ and $p_{switch} = 0.0$ here $\Delta = 1.964 \pm 0.030$.}
			\label{Cluster}
		\end{center}
	\end{figure}
	
	
	\begin{figure}[h!]
		\begin{center}
			\hspace*{-0.1in}
			\includegraphics[angle=-90, trim = 0 0 0 0, clip = true, width=0.75\linewidth]{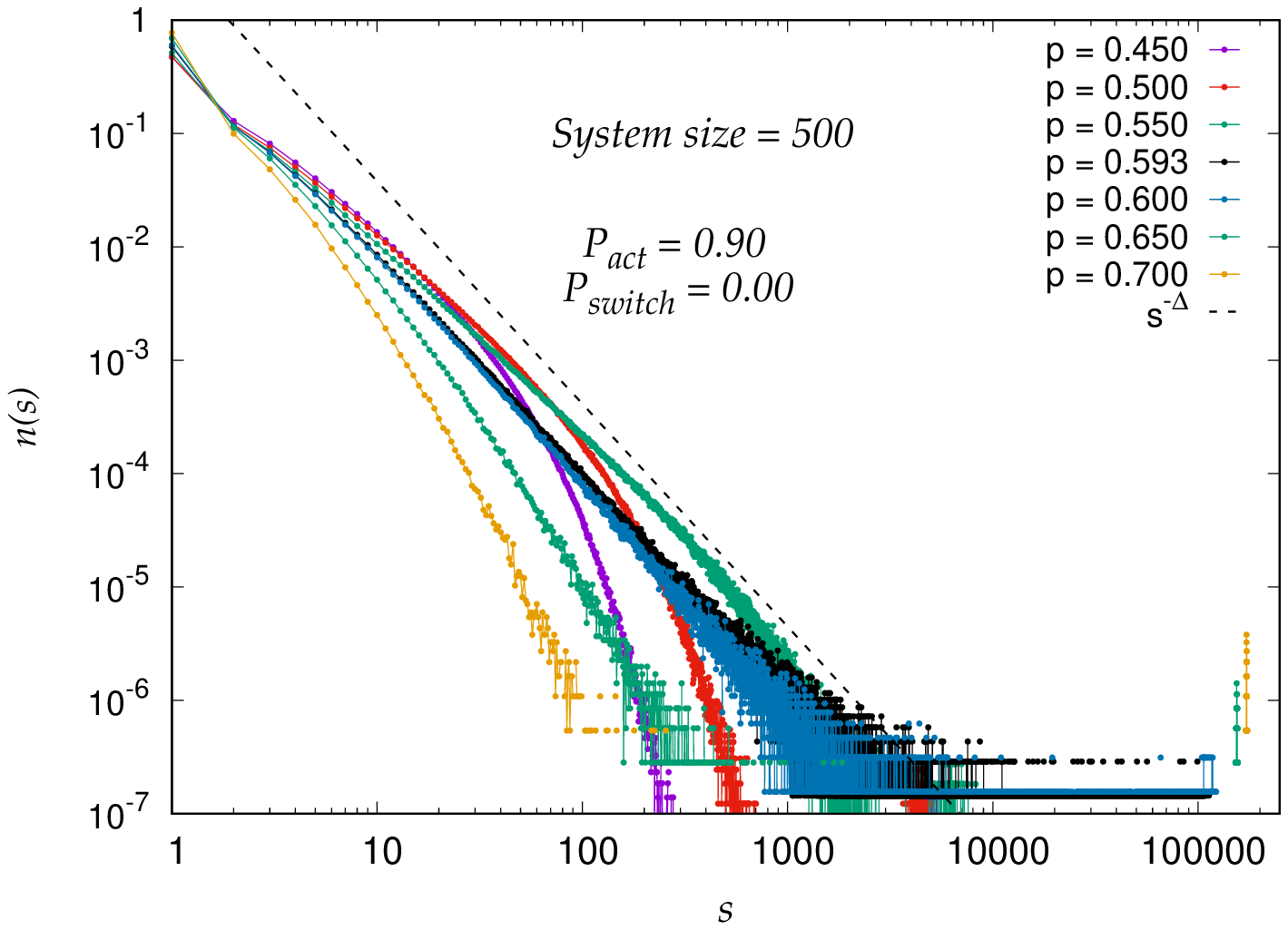}
			\caption{$n(s)$ versus $s$ with $L = 500$, $p_{act} = 0.90$ and $p_{switch} = 0.0$ here $\Delta = 1.911 \pm 0.002$.}
			\label{ClusterPact0.90}
		\end{center}
	\end{figure}
	
	The cluster distribution for $p_{act}=1.0$ and $p_{switch}=0.0$ for the cluster of inactive sites is shown in Fig. \ref{Cluster}. Here HK algorithm has been used to study \cite{Kopelman} the cluster distribution. Here $\Delta$ is found to be close to $1.9$. The distribution is shown not only for $p_c$ but also above and below $p_c$. This nicely agrees with \cite{Mario}. In Fig. \ref{ClusterPact0.90} the cluster distribution is shown for $p_{act}=0.9$ and $p_{switch}=0.0$.


	\begin{figure}[h!]
		\begin{center}
			\hspace*{-0.1in}
			\includegraphics[angle=-90, trim = 0 0 0 0, clip = true, width=0.75\linewidth]{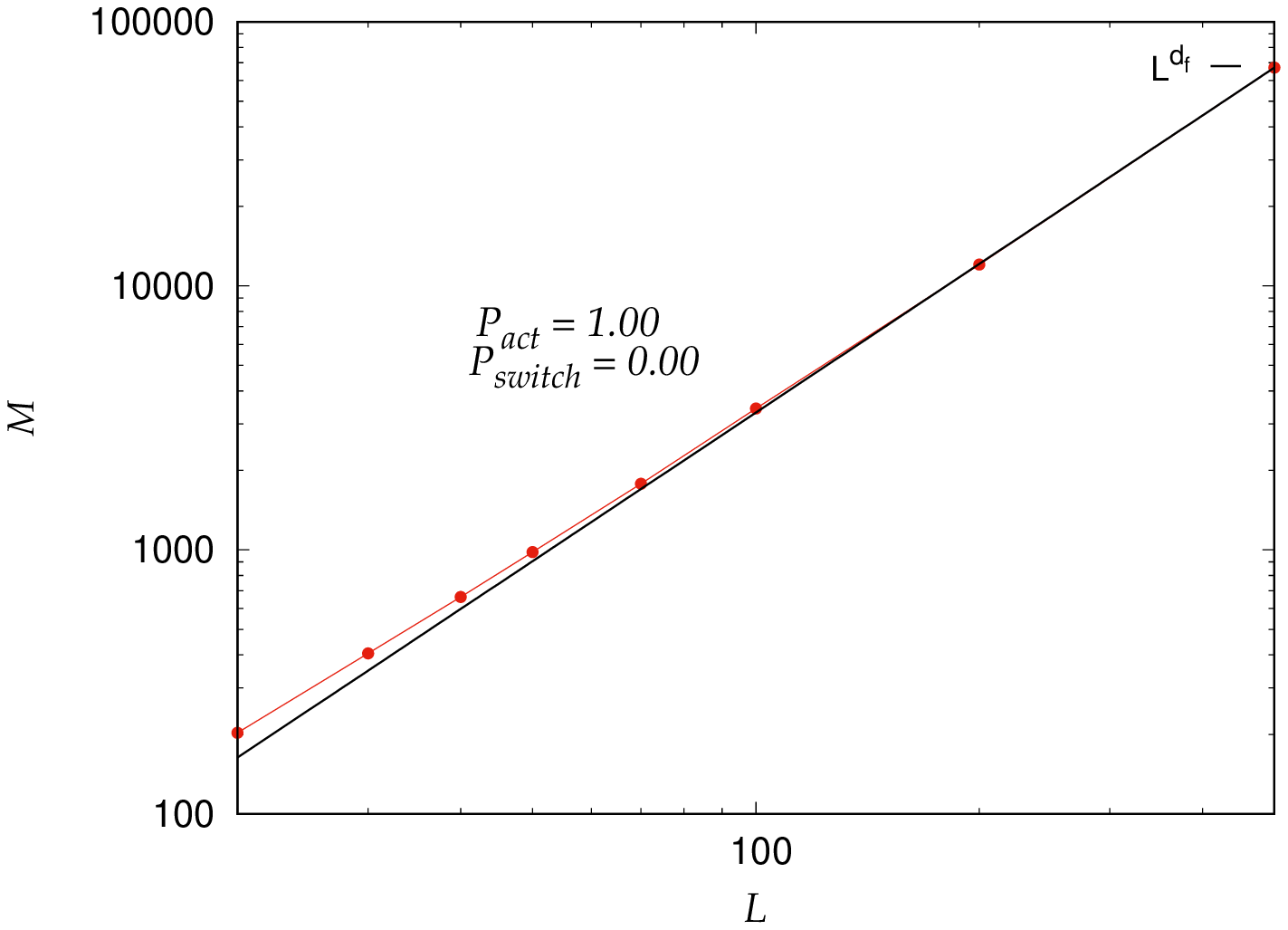}
			\caption{$M$ versus $L$ with  $p_{act} = 1.00$ and $p_{switch} = 0.0$ here $d_f = 1.811 \pm 0.011$.}
			\label{M_L}
		\end{center}
	\end{figure}
	
	\begin{figure}[h!]
		\begin{center}
			\hspace*{-0.1in}
			\includegraphics[angle=-90, trim = 250 0 0 0, clip = true, width=0.99\linewidth]{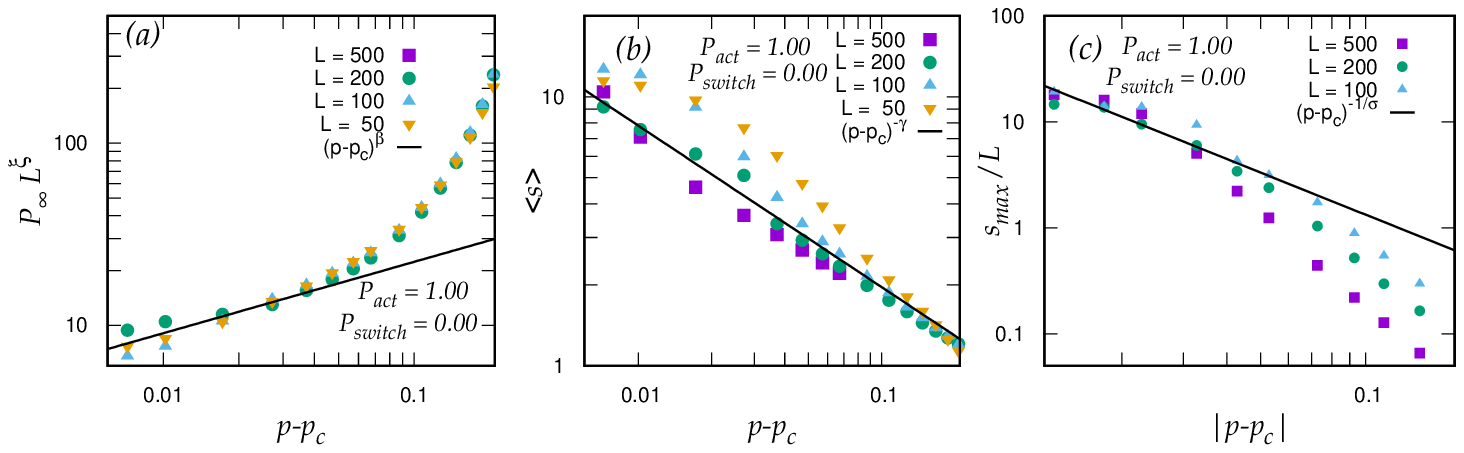}
			\caption{Study of critical exponents : (a) $P_{\infty}L^{\xi}$ versus $p-p_c$ with $\bar{\beta} = 0.392 \pm 0.045$ and $\xi = 1.85$, which gives $\beta = 2.859$, (b) $\braket{s}$ versus $p-p_c$ with $\gamma = 0.600 \pm 0.021 $ and (c) $\frac{s_{max}}{L}$ versus $|p-p_c|$ with $\bar{\sigma} = 0.757 \pm 0.150$, which gives $\sigma = 0.377$. These exponents approximately satisfy the relation $\beta + \gamma = \frac{1}{\sigma}$. }
			\label{critical_exponents}
		\end{center}
	\end{figure}
	
	Mass of the largest cluster $M$ for $p=p_c$ shows a typical behaviours $M \sim L^{d_f}$ where $L$ is the system size and $d_f$ is the fractal dimension. This is shown in Fig. \ref{M_L}. Here $d_f$ is  found to be $d_f=1.811 \pm 0.011$. 
	
	From the values of system dimension $d$ and fractal dimension $d_f$ we may find $\Delta$ from the hyperscaling relation $\Delta=\frac{d}{d_f}+1$. The value of $\Delta$ thus obtained is close to $2.10$. However, as mentioned earlier, we have obtained a smaller value for $\Delta$. 
	
	The strength of the infinite cluster $P_{\infty}$ and the divergence of the mean cluster size $\langle s \rangle$ are another two important quantities. Those result in two more exponents $\beta$ and $\gamma$ which for our case happens to be $2.859$ and $0.600 \pm 0.021 $ and are respectively shown in Fig. \ref{critical_exponents} (a) and (b).
	
	From Fig. \ref{Cluster} another quantity is studied very often which is the cut-off cluster size $s_{max}$. This can be found from Fig. \ref{Cluster} (c) by using the well known form of $n_s \sim s^{-\Delta}\exp(-s/s_{max})$ for large $s$. The corresponding variation of cut-off cluster size $s_{max}$ for our system is shown in Fig. \ref{critical_exponents}. From this we found the exponent $\sigma$ as $0.377$, the relation being $s_{max} \propto (p-p_c)^{-1/\sigma}$.   
	
	The value of the exponent $\alpha$ can be found out easily as $\sum_0^\infty n_s = (p-p_c)^{2-\alpha}$. The value of $\alpha$ obtained in our case is  close to $2.0004$. 
	
	It is clear from the above facts that the Rushbrooke's inequality $\alpha+2\beta+\gamma \geq 2$ is satisfied. Also the relation $\beta+\gamma = \frac{1}{\sigma}$ is verified here to be approximately true.

	\section*{Summary and Discussions}
	In summary, the work may be considered to have two aspects. Firstly, we tried to model the heart based on percolation theory and have studied several related features. Secondly, we have extracted the relevant exponents from the model so obtained.  
	
	In the first part, as mentioned, the signal propagation through the heart have been studied. It is known that the structure as well as the physiological processes in the heart are the key factors depending on which the signal propagates. We treat the heart as a 2D square grid of $L \times L$ cells which may be ``waiting (0)",  ``active(1)" and ``inactive(2)". We started from some initial distribution of $0$s and $2$s. As the signal (AP) propagates, there are  $1$s (active sites) too. We considered two varieties: 1. When $2\rightarrow0$ transition is allowed, i.e., $p_{switch} \neq 0$ and $p_{act}=1$; 2. When $2\rightarrow0$ transition is not allowed, i.e., $p_{switch} = 0$ and $p_{act} \neq 1$. We studied several important quantities. The number of arrival $N_A$ at the other side of the grid shows a Gaussian behaviour when $p_{act}$ is varying and $p_{switch}=0.0$. This indicates a kind of inhibitory behaviour and the AP/signal simply passes down the system within a fixed time. After that no new signal can pass down through the system as the cells are in their refractory period for an infinitely long time, indicating a damaged heart. However, the number of arrival $N_A$ at the other side of the grid shows a damped oscillatory behaviour when $p_{act}=1.0$ and $p_{switch}$ is varying, saturating to a finite value for high $p_{switch}$. This indicates that there are always a finite number of active cells and therefore new APs can pass through the system. It is to be noted that very high value of $p_{switch}$ has a disadvantage as the AP is trapped within the system which may be a cause for irregular heart rhythms. 
	
	In the next part of the work, we have studied the percolation model of the system and analyzed the critical exponents for the same. It has been observed that the critical probability for percolation $p_c$ is close to $0.5925$ for the $p_{switch} = 0$ and $p_{act} = 1$. 
	\begin{table}[h!]
		\begin{center}
			\begin{tabularx}{0.4\textwidth}{
                                            |>{\centering\arraybackslash}X
                                            |>{\centering\arraybackslash}X
                                            |>{\centering\arraybackslash}X|}
				\hline
				\hspace{0.0cm}$p_{act}$\hspace{0.0cm}  & \hspace{0.0cm}$p_{switch}$\hspace{0.0cm}&  \hspace{0.0cm}$p_{c}$\hspace{0.0cm}\\  
				\hline
				\hline
				1.00	&	0.00   &	0.5925\\
				\hline
				0.95   &   0.00    &   0.6185\\
				\hline
				0.90   &   0.00    &   0.6476\\
				\hline
				0.80   &   0.00    &   0.7157\\
				\hline
				1.00   &   0.03    &   0.12\\
				\hline
			\end{tabularx}
		\end{center}
		\caption{Approximate values of $p_c$ for different $p_{act}$ and $p_{switch}$}
		\label{pc_table}
	\end{table}
	The value of $p_c$ shifts towards higher and higher side as $p_{act}$ is made smaller and smaller. Keeping $p_{act}$ fixed at $1$ and changing $p_{switch}$ a little has severe effect on $p_c$ and it shifts abruptly towards $0$. The way $p_c$ is changing is shown in Table \ref{pc_table}. 
	
	\begin{table}[h!]
		\begin{center}
			\begin{tabular}{| c | c | c | c | c | c | c |}
				\hline
				Model & $\Delta$ & $\alpha$ & $\beta$ & $\gamma$ & $\sigma$  &  $\nu$ \\  
				\hline
				\hline
				Heart Model   &	    &     &   &    & 	&   \\
				$p_{act}=1.0$ &	$1.964$    &  $2.0004$ & $2.859$ & $0.600$ & $0.377$   &  $4/3$ \\
				$p_{switch}=0.0$ & & & & & &\\
				\hline
				Standard Model & $187/91$ & $-2/3$ & $5/36$ & $43/18$ & $36/81$  & $4/3$\\
				\hline
			\end{tabular}
		\end{center}
		\caption{Comparison of critical exponents.}
		\label{exp_table}
	\end{table}
	
	The tortuosity is studied and the related exponent $u \simeq 0.205$ is obtained.
	We also studied the exponents relevant to percolation theory. In this part, we considered the parameters $p_{switch} = 0$ and $p_{act} = 1$ only. The cluster distribution is studied here as shown in Fig. \ref{Cluster}. From this, we checked the exponents $\alpha, \beta, \gamma, \sigma$ etc. which are shown in Table \ref{exp_table}. It has been observed that a few exponents are quite close to those obtained in case of standard percolation. However, other exponent values deviate from the standard case. This may be due to the transformation of cells as $0 \rightarrow 1 \rightarrow 2 \rightarrow 0$, which is a system specific property and therefore is not exactly similar to the standard percolation. The Rushbrooke Inequality $\alpha+2\beta+\gamma \geq 2$ and the relation $\beta+\gamma = 1/\sigma$ has been observed to be closely satisfied.

	\begin{center}
		\textbf{Acknowledgement}
	\end{center}
	
	AM acknowledges financial support from CSIR (SRF Grant no. 08/0463(12870)/2021-EMR-I). AM and SG acknowledges the computational facility of Vidyasagar College.

\end{document}